\documentclass[manuscript]{acmart}
\AtBeginDocument{%
  }
\setcopyright{acmlicensed}
\copyrightyear{2026}
\acmYear{2026}
\acmDOI{XXXXXXX.XXXXXXX}
\usepackage{xspace}
\usepackage{float}
\usepackage{textcomp}
\begin{document}
\newcommand{\Sys}{\textsc{AnchorNote}\xspace}
\title[\Sys{}]{\Sys{}: Exploring Speech-Driven Spatial Externalization for Co-Located Collaboration in Augmented Reality}

\author{Diya Hundiwala}
\email{diyahundiwala@princeton.edu}
\affiliation{
  \institution{Princeton University}  \city{Princeton}
  \state{NJ}
  \country{USA}
}

\author{Andrés Monroy-Hernández}
\email{andresmh@cs.princeton.edu}
\affiliation{
  \institution{Princeton University}
  \city{Princeton}
  \state{NJ}
  \country{USA}
} 
\begin{abstract}
Sticky notes remain a durable collaborative medium because they support rapid idea externalization, rearrangement, and coordination of group attention through spatial organization while being low-friction and lightweight \cite{Ball2021StickyNotes, Subramonyam2019AffinityLens}. Recent AR systems suggest new ways to externalize ideas in shared physical space, including spatial annotations and digital workspaces \cite{Miyazaki2025JustTalkStickyNotes, Jensen2018Remediating, StickyAR2017, StickyNotesAR2019, Zhang2024AiRNote}. We introduce \Sys{}, a co-located AR system that lets collaborators intentionally capture spoken ideas as spatially anchored sticky notes via live transcription and LLM summarization. We evaluated \Sys{} in a two-phase iterative study with 20 participants completing a brainstorming and thematic grouping task to examine how speech-driven, spatially persistent capture shapes idea externalization in collaboration. We found that \Sys{} reduced writing effort but reshaped collaboration by introducing new coordination costs and shifting how participants formulated, timed, and organized ideas. We use \Sys{} as an exploratory probe to study how speech-driven, spatial externalization in AR restructures collaborative cognition and coordination, and to derive design implications for future co-located AR collaboration tools.
\end{abstract}

\begin{teaserfigure}
   \includegraphics[width=1\textwidth]{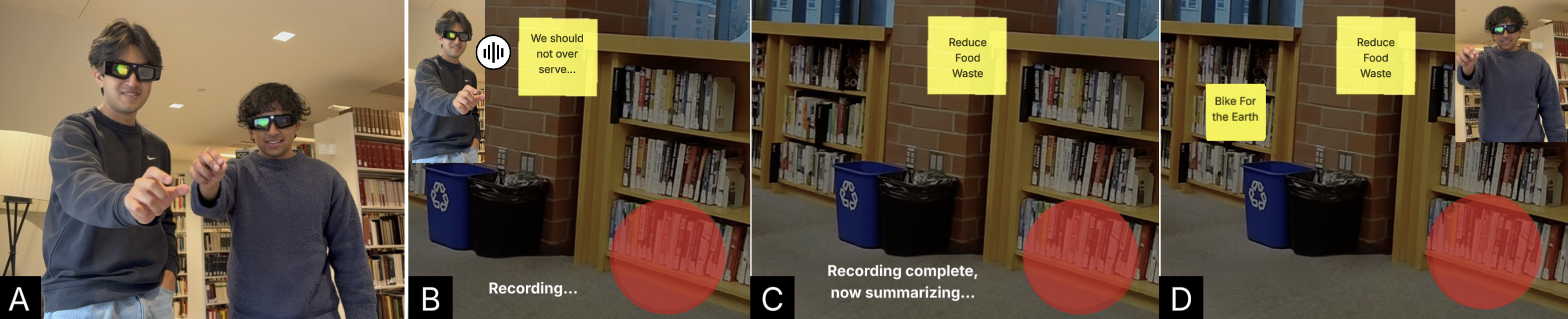}
   \caption{\Sys{} end-to-end workflow: (A) Users prepare to use \Sys side by side. (B) User 1 starts a note via the red button and \Sys transcribes speech live. (C) User 1 stops recording and \Sys summarizes into a short title. (D) Another session participant sees User 1's note and creates another, resulting in two shared notes anchored in the same physical workspace across devices.}
   \Description[Screenshots of users interacting with the system]{The figure shows users interacting with an Augmented Reality system for speech-driven sticky notes.}
   \label{fig:teaser}
\end{teaserfigure}
\maketitle



\section{INTRODUCTION}

Groups brainstorm by weaving together conversation and shared attention, often mediated through externalized artifacts \cite{dourish1992awareness, heath1991collaborative}. Sticky notes are a widely used medium for this work: they support spatial externalization and iterative grouping, helping teams maintain awareness of contributions and relationships among ideas \cite{Ball2021StickyNotes, fischelhalskovstickynotes}. However, spoken contributions remain fragile in these settings. If no one explicitly captures them, ideas voiced in conversation can quickly disappear, and responsibility for accurate documentation often falls unevenly on whoever is writing the notes \cite{crosstalk, hindus1993capturing, Jensen2018Remediating, Miyazaki2025JustTalkStickyNotes, costley2021collaborative}.

Augmented reality (AR) presents a new medium to externalize ideas in a shared collaborative space; prior systems have explored digital sticky notes via large tablet interfaces and automated speech capture \cite{Miyazaki2025JustTalkStickyNotes, Jensen2018Remediating, StickyAR2017, StickyNotesAR2019, Zhang2024AiRNote}. To investigate what is gained, lost, and transformed when externalization becomes speech-driven and spatially persistent on AR glasses, we present \Sys{}, a co-located sticky-note system that lets collaborators capture spoken ideas as shared, spatially anchored notes. Unlike prior AR sticky-note and speech-capture systems that emphasize automation or single-user productivity, \Sys{} focuses on intentional speech capture during live, face-to-face collaboration where turn-taking and shared attention are central. 

We studied \Sys{} through a two-phase iterative investigation with 20 participants working in pairs on a brainstorming and thematic grouping task. Phase~1 surfaced breakdowns and coordination costs relative to analog sticky notes, while Phase~2 examined how targeted interaction changes repaired those breakdowns. Our findings map tradeoffs in speech-driven spatial externalization, including shifts from writing burden to system monitoring, the conditional benefits of spatial persistence, and how interaction legibility introduces shifts in conversational coordination. We do not argue that AR sticky notes replace analog practices; instead, we use \Sys{} as an exploratory system to examine how speech-driven, spatial capture within AR restructures collaborative dynamics.

\section{RELATED WORK}

\subsection{Sticky Notes and Spatial Externalization in Collaborative Ideation}
Externalizing ideas onto physical artifacts supports memory offloading and collective reasoning \cite{Ball2021StickyNotes, kirsh2010thinking, zhang1994representations}. Sticky notes are particularly effective because they enable rapid capture and spatial rearrangement, helping groups surface relationships among ideas and iteratively refine structure \cite{Ball2021StickyNotes, fischelhalskovstickynotes}. However, paper sticky notes often interrupt conversation and can constrain how much spoken detail is documented when teams must write brief fragments mid-discussion \cite{Jensen2018Remediating, costley2021collaborative, Miyazaki2025JustTalkStickyNotes}. This motivates examining how alternative capture modalities (e.g., speech-driven capture) change the costs and coordination dynamics of externalization during live collaboration.

\subsection{AR Workspaces for Co-Located Collaboration and Spatial Annotation}
Co-located AR can extend shared workspaces by anchoring digital content to physical space, supporting joint attention and shared context \cite{dynamicland, billinghurst1999collaborative, yang2024fostering}. Taken together, prior AR workspace and annotation systems demonstrate the promise of spatial organization, but they also surface recurring challenges in alignment, shared attention, and breakdown recovery; at the same time, speech-driven capture systems reduce manual input yet often prioritize automation or are evaluated in screen-based collaborative settings \cite{luo2025documents, Subramonyam2019AffinityLens, ens2021, Miyazaki2025JustTalkStickyNotes, Zhang2024AiRNote, StickyAR2017, StickyNotesAR2019}. This leaves limited evidence about how intentional speech capture into shared, spatially persistent artifacts affects turn-taking and coordination during live, co-located collaboration on head-worn AR.

\subsection{Embodied and Speech-Driven Interaction for Collaborative Capture}
Gesture-based interaction can support hands-free manipulation, but in dynamic settings it often suffers from ambiguity and false activations that increase cognitive overhead \cite{rico2010usable}. Speech-driven capture and transcription can reduce the burden of manual input, but prior systems frequently prioritize automation over user control or evaluate collaboration in digital whiteboard contexts rather than co-located embodied AR \cite{Zhang2024AiRNote, Miyazaki2025JustTalkStickyNotes, Jensen2018Remediating}. Taken together, prior research leaves an open question: \textit{how can co-located AR systems combine speech-driven capture with spatial persistence while maintaining user agency and preserving conversational flow?} We use \Sys{} as an exploratory probe to examine this question in head-worn, co-located AR, where timing, attention, and breakdown recovery can directly alter group coordination.

\section{THE \Sys{} SYSTEM}
\begin{figure*}
    \centering\includegraphics[width=0.85\linewidth]{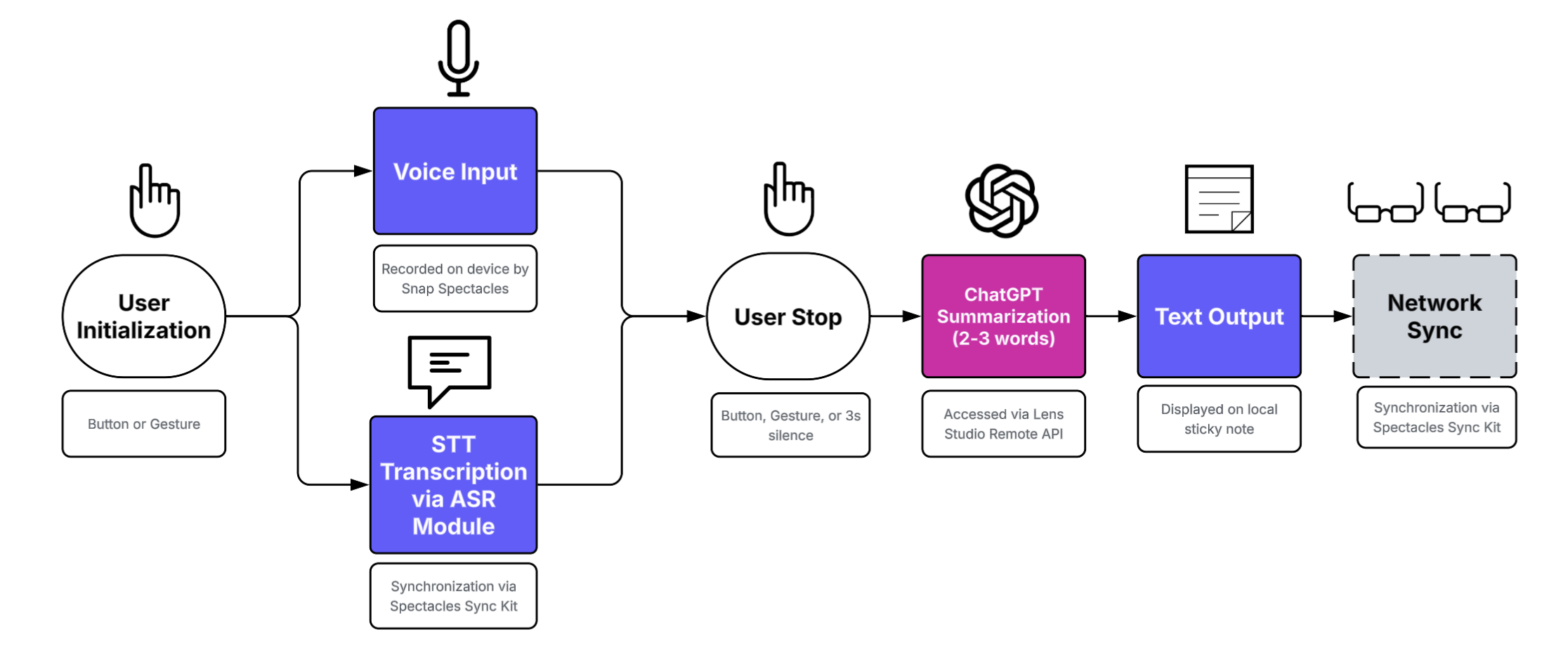}
    \caption{High-level interaction pipeline for shared speech-to-note capture in \Sys{} (details in Appendix~\ref{appendixImplementation}).}
    \label{fig:gameflow}
\end{figure*}

\Sys{} enables multiple users wearing AR glasses to create, view, and organize shared sticky notes in a common physical space. In a typical session, collaborators stand in a shared environment and use a wall or open space as a brainstorming surface. Users start and stop capture via an explicit control (gesture in Phase~1; button in Phase~2). During capture, the system displays live transcription on a sticky note appearing in the user's view at a fixed distance (50 \,cm) and a clear recording indicator; after capture, the system summarizes the transcript into a short title that appears on the note using an LLM (Appendix~\ref{appendixPrompt}). The system spatially anchors notes and shares them across devices, allowing collaborators to reference and reorganize them together in the same physical space. Figure~\ref{fig:gameflow} provides a high-level view of the interaction pipeline and shared-note synchronization. We provide implementation details (platform components, fallback behavior, and synchronization mechanisms) in Appendix~\ref{appendixImplementation}.

\section{EVALUATION}
We conducted a two-phase iterative design investigation of \Sys{} in co-located collaboration. Phase~1 was diagnostic, surfacing breakdowns, coordination costs, and attention shifts introduced by speech-driven AR note capture. Phase~2 was reparative, examining how targeted interaction changes (explicit controls, clearer system-state signaling, and lightweight curation) altered those breakdowns. Across both phases, pairs brainstormed campus-related topics and grouped ideas spatially. We audio- and video-recorded the sessions. Our university's IRB approved all procedures.

\subsection{Participants and Recruitment}
We recruited 20 undergraduate students (ages 18-22) from a private university via word of mouth. Participants reported varied prior familiarity with AR. We compensated participants \$20 for a 30--40 minute session. Phase~1 included 12 participants; four Phase~1 pairs returned for Phase~2 (8 participants).

\subsection{Procedure}
\begin{figure*}
    \centering\includegraphics[width=0.8\linewidth]{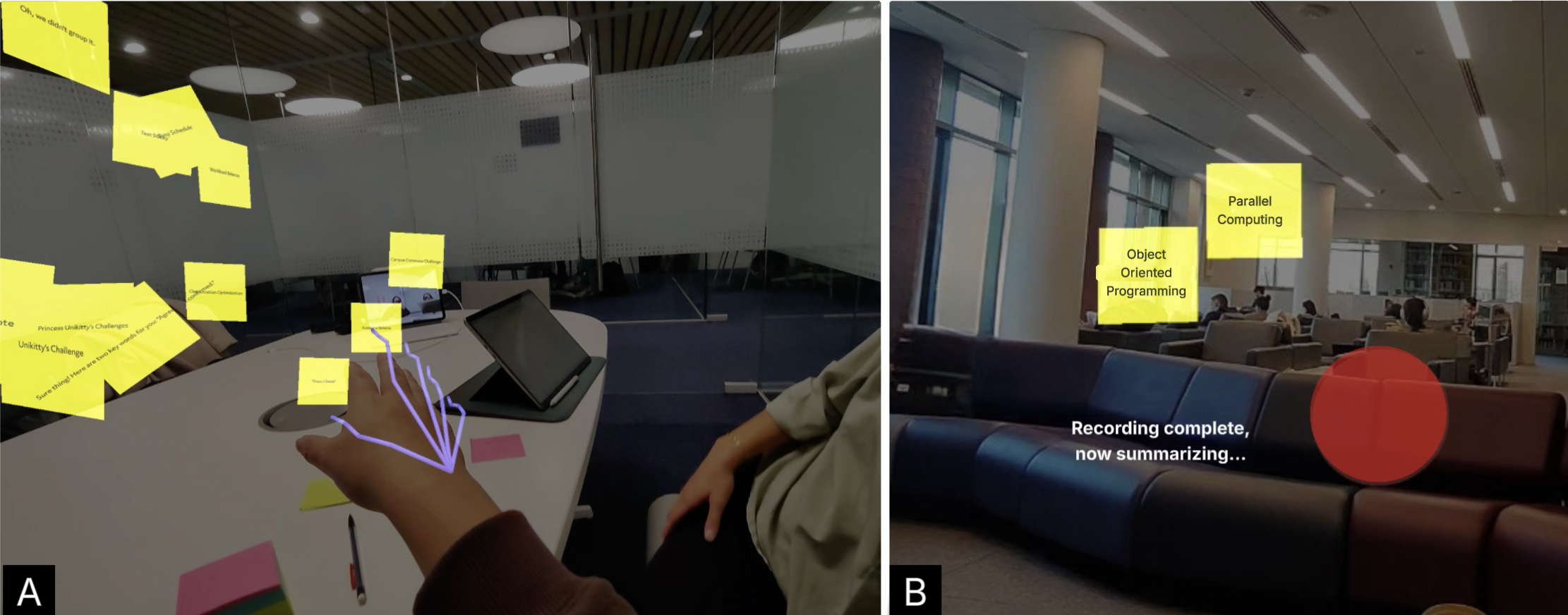}
    \caption{(A) System design evaluated in Phase 1 with gesture-triggered note creation, no deletion mechanism, and no system state indicators. (B) System design evaluated in Phase 2 with button-triggered note creation, a "hold-to-delete" functionality, and system state indicators such as "Recording" and "Recording complete, now summarizing...".}
    \label{fig:sysdesigncomparison}
\end{figure*}

In Phase~1, pairs completed two 10--12 minute brainstorming sessions: one using analog sticky notes and one using \Sys{} (counterbalanced order). Participants first completed a brief demographic and AR familiarity questionnaire (Appendix~\ref{appendixIntro}) and received a walkthrough of \Sys{} with 3--5 minutes of practice. Then, participants engaged in brainstorming, completing a post-task survey measuring conversational flow, cognitive effort, coordination, and perceived collaboration (Appendix~\ref{appendixSurvey}) after each condition. Sessions concluded with a semi-structured interview (Appendix~\ref{appendixInterview}). Early Phase~1 sessions seated participants across from one another; later sessions positioned pairs side-by-side to reduce physical interference during embodied interaction. In Phase~2, returning pairs completed a 15-minute brainstorming session with the same prompt structure using an updated \Sys{} version with button note-creation control, clearer system-state indicators, and a delete mechanism (Figure~\ref{fig:sysdesigncomparison}). Participants then completed the same survey and an interview focused on perceived differences relative to Phase~1.

\subsection{Analysis}
We use quantitative survey responses as supporting signals to contextualize participants' accounts. We report descriptive trends across the analog baseline, Phase~1, and Phase~2. Given small sample sizes and the iterative design, we report descriptive statistics to contextualize qualitative accounts rather than conducting inferential tests. We analyzed interviews and session recordings using thematic analysis, coding for breakdowns, recovery strategies, attention shifts, coordination around capture, and how spatial notes shaped organization over time.  We lightly edited quotes for clarity while preserving participants' meaning.

\section{RESULTS}
\subsection{Immediate Speech-to-Note Externalization Reshaped Ideation}
A distinctive effect of speech-driven capture was that externalization became immediate and public: as participants spoke, \Sys{} converted their words into a visible note. Several participants described adapting their ideation strategies in response, reporting pressure to internally finalize an idea before speaking because it would be ``recorded'' as an artifact. As P10 explained, \textit{"when writing on [analog sticky notes], it felt like I could think through ideas before writing it down. [With AR], it felt like I had to have the idea finalized."} This shift sometimes suppressed tentative contributions and altered conversational rhythm, especially when participants were uncertain about when the system was listening or how the summary would represent nuance. Together, these accounts suggest that speech-driven capture does not simply remove friction; it can restructure cognitive flow by changing when ideas feel ``ready'' to share.

\subsection{Writing Burden Decreased, but Monitoring and Recovery Work Increased}
Participants often valued not needing to interrupt discussion to handwrite notes (e.g., P3 noted they were \textit{"no longer writing while thinking so could just think out loud"}). However, Phase~1 frequently shifted effort from writing to monitoring whether the system had correctly triggered, recorded, and transcribed speech. P9 described being \textit{"focused on the system rather than the prompt."} When mistranscriptions occurred, the lack of direct editability increased recovery costs: correcting errors required deleting and re-creating notes, sometimes repeating ideas and further interrupting flow. Participants also expressed concern about losing nuance through summarization; as P10 noted, \textit{"When I write things down, I know what exactly to emphasize, but I don't know that the AI can capture the nuances."} These accounts aligned with survey trends indicating higher perceived effort and interruptions in Phase~1 relative to analog (Appendix~\ref{appendixQuantTable}), reinforcing that supervision and recovery were central sources of cost.

\subsection{Interaction Ambiguity Disrupted Conversational Timing; Explicit Controls Repaired Coordination}
In Phase~1, gesture misfires and transcription delays produced breakdowns that affected turn-taking and conversational timing. Participants paused mid-thought, repeated ideas, or coordinated explicitly about when to capture notes. P10 reflected that \textit{"the system became the center of the conversation, rather than being a side tool."} Gesture triggering in particular introduced ambiguity about system state and accidental activation, which amplified monitoring behavior and made capture feel socially disruptive. P1 noted that gesture \textit{``triggers were confusing,''} and described a mistranscription (e.g., ``open sesame'' when the intended phrase was ``too much work'') that derailed the conversation and required explicit recovery. Such errors often forced participants to pause, repeat ideas, or negotiate whether to delete and re-capture notes, further disrupting conversational timing. In Phase~2, explicit button-based note creation and clearer system-state indicators reduced accidental capture and made the interaction boundaries more legible. P5 emphasized that the system \textit{"was a lot easier to use when it was tied to one button and told you when something was being transcribed versus summarizing."} Although the interaction was less ``hands-free,'' participants described it as easier to coordinate around, helping capture recede into the background of collaboration.

\subsection{Spatial Persistence Supported Shared Reference, but Clutter Quickly Undermined Sensemaking Without Curation}
Across both phases, participants described spatially anchored notes as useful for shared reference and organization (e.g., P3 noted that \textit{"being able to see [ideas] in 3D space helped."}). However, in Phase~1 the absence of deletion mechanisms led to clutter that undermined grouping and increased coordination overhead. As P4 described, \textit{"too many stickies [were] being generated,"} making organization more difficult over time. In Phase~2, adding lightweight curation (delete) enabled participants to repair mistakes and keep the workspace legible, sustaining the benefits of spatial persistence. As P9 explained, AR made it \textit{"much easier to organize thoughts... if you didn't like it you could delete it."}. 

We report additional qualitative observations in Appendix~\ref{appendixQual}.

\section{DISCUSSION}
Our findings suggest that speech-based spatial capture externalization reshapes collaboration by shifting effort, altering conversational timing, and changing when ideas feel ``ready'' to share. Rather than viewing these effects as implementation flaws alone, we interpret them as design tensions that future co-located AR tools must navigate.
Immediate conversion of speech into an artifact increased pressure to pre-formulate thoughts before speaking, which could suppress tentative ideation and alter conversational dynamics. This suggests that speech-based capture is not a neutral input channel: by moving externalization earlier in the thinking process, it can reshape how ideas are generated and shared. Designers may mitigate this by supporting provisionality (e.g., draft states, easy revision, or delayed publishing) and by making capture boundaries explicit.
Although participants valued reduced manual writing, ambiguous system state and error recovery demands became the dominant source of burden in Phase~1. Gesture triggering felt expressive but introduced ambiguity that disrupted conversational timing and required explicit coordination, while explicit controls and system-state cues improved predictability and made capture easier to coordinate around, even at the cost of ``naturalness.'' Together, these findings suggest that for co-located AR tools, interaction legibility and fast recovery matter more than hands-free novelty. Spatial persistence supported shared understanding only when workspaces remained editable; without delete functionality, clutter quickly undermined sensemaking.

\subsection{Limitations and Ethics/Accessibility Considerations}
Our study involves a small sample of undergraduate participants and a prototype constrained by current AR-hardware capabilities; observed breakdowns may differ with other devices and settings. Because Phase~2 involved returning participants, improvements may partly reflect familiarity in addition to the Phase~2 design changes. Reliance on AR hardware limits access, and gesture-based interaction may disadvantage users with motor impairments. Speech transcription and LLM summarization raise privacy and transparency concerns.

\section{LLM DISCLOSURE}
We used ChatGPT (GPT-5.2) to assist with drafting, editing, and code support within this paper, as well as during system development for debugging and study design brainstorming. We used the Lens Studio MCP Server in Cursor as well during system development.

\section{CONCLUSION}
We presented \Sys{}, a co-located AR sticky-note prototype that intentionally captures spoken contributions into spatially anchored notes. Through a two-phase diagnostic-and-repair investigation, we mapped tradeoffs in speech-driven spatial externalization: manual writing burden decreased, but monitoring and recovery work increased when system state was ambiguous; spatial persistence supported shared reference yet clutter undermined sensemaking without deletion; and gesture triggering introduced coordination costs that explicit controls helped repair. Notably, immediate speech-to-note externalization reshaped ideation strategies by encouraging pre-formulation of ideas before speaking.
We do not argue that AR sticky notes are ready to replace analog practices. Instead, we use \Sys{} to highlight how capture modality and interaction legibility can restructure collaborative cognition and coordination, and to inform the design of future co-located AR tools that better support provisionality, fast recovery, and workspace curation. Future work should explore richer multi-speaker support, more accessible input modalities, voice note storage, and interaction techniques that prioritize robustness and fast error recovery in co-located AR collaboration.
\newpage
\bibliographystyle{ACM-Reference-Format}
\bibliography{references}

@article{Ball2021StickyNotes,
  author  = {Ball, Linden J. and Christensen, Bo T. and Halskov, Kim},
  title   = {Sticky notes as a kind of design material: How sticky notes support design cognition and design collaboration},
  journal = {Design Studies},
  volume  = {76},
  year    = {2021},
  pages   = {101034},
  issn    = {0142-694X},
  doi     = {10.1016/j.destud.2021.101034},
  url     = {https://www.sciencedirect.com/science/article/pii/S0142694X21000454}
}

@inproceedings{Subramonyam2019AffinityLens,
  author    = {Subramonyam, Hariharan and Drucker, Steven M. and Adar, Eytan},
  title     = {Affinity lens: Data-assisted affinity diagramming with augmented reality},
  booktitle = {Proceedings of the 2019 CHI Conference on Human Factors in Computing Systems},
  series    = {CHI '19},
  year      = {2019},
  publisher = {Association for Computing Machinery},
  address   = {New York, NY, USA},
  location  = {Glasgow, Scotland, UK},
  pages     = {1--13},
  numpages  = {13},
  doi       = {10.1145/3290605.3300628},
  url       = {https://doi.org/10.1145/3290605.3300628}
}

@inproceedings{Jensen2018Remediating,
  author    = {Jensen, Mads M{\o}ller and R{\"a}dle, Roman and Klokmose, Clemens N. and B{\o}dker, Susanne},
  title     = {Remediating a design tool: Implications of digitizing sticky notes},
  booktitle = {Proceedings of the 2018 CHI Conference on Human Factors in Computing Systems},
  series    = {CHI '18},
  year      = {2018},
  publisher = {Association for Computing Machinery},
  address   = {New York, NY, USA},
  location  = {Montr{\'e}al, QC, Canada},
  articleno = {224},
  numpages  = {12},
  doi       = {10.1145/3173574.3173798},
  url       = {https://doi.org/10.1145/3173574.3173798}
}

@inproceedings{Miyazaki2025JustTalkStickyNotes,
  author    = {Lee, Gippeum and Park, Namchoon},
  title     = {Just talk, and sticky notes will be created: Towards collaborative dialogue in face-to-face workshops with a generative {AI}-based tool},
  booktitle = {Proceedings of the Extended Abstracts of the CHI Conference on Human Factors in Computing Systems},
  series    = {CHI EA '25},
  year      = {2025},
  publisher = {Association for Computing Machinery},
  address   = {New York, NY, USA},
  articleno = {363},
  numpages  = {6},
  doi       = {10.1145/3706599.3720231},
  url       = {https://doi.org/10.1145/3706599.3720231}
}

@misc{StickyAR2017,
  author       = {{Uriverse, Inc.}},
  title        = {Sticky – {AR} sticky notes},
  year         = {2017},
  howpublished = {\url{https://apps.apple.com/us/app/sticky-ar-sticky-notes/id1109175965}},
  note         = {iOS app. Accessed: 2026-01-08}
}

@misc{StickyNotesAR2019,
  author       = {{Cunum Games}},
  title        = {Sticky notes {AR}},
  year         = {2019},
  howpublished = {\url{https://play.google.com/store/apps/details?id=com.cunum.stickynotesar}},
  note         = {Google Play app. Accessed: 2026-01-05}
}

@inproceedings{dourish1992awareness,
  author    = {Dourish, Paul and Bellotti, Victoria},
  title     = {Awareness and coordination in shared workspaces},
  booktitle = {Proceedings of the 1992 ACM Conference on Computer-Supported Cooperative Work},
  series    = {CSCW '92},
  year      = {1992},
  publisher = {Association for Computing Machinery},
  address   = {New York, NY, USA},
  location  = {Toronto, Ontario, Canada},
  pages     = {107--114},
  numpages  = {8},
  doi       = {10.1145/143457.143468},
  url       = {https://doi.org/10.1145/143457.143468}
}

@inproceedings{heath1991collaborative,
  author    = {Heath, Christian and Luff, Paul},
  title     = {Collaborative activity and technological design: Task coordination in London Underground control rooms},
  booktitle = {Proceedings of the Second European Conference on Computer-Supported Cooperative Work (ECSCW '91)},
  year      = {1991},
  publisher = {Springer Netherlands},
  address   = {Dordrecht, The Netherlands},
  location  = {Amsterdam, The Netherlands},
  pages     = {65--80},
  doi       = {10.1007/978-94-011-3506-1_5},
  url       = {https://doi.org/10.1007/978-94-011-3506-1_5}
}

@inproceedings{crosstalk,
  author    = {Xia, Haijun and Wang, Tony and Gunturu, Aditya and Jiang, Peiling and Duan, William and Yao, Xiaoshuo},
  title     = {CrossTalk: Intelligent substrates for language-oriented interaction in video-based communication and collaboration},
  booktitle = {Proceedings of the 36th Annual ACM Symposium on User Interface Software and Technology},
  series    = {UIST '23},
  year      = {2023},
  publisher = {Association for Computing Machinery},
  address   = {New York, NY, USA},
  location  = {San Francisco, CA, USA},
  articleno = {60},
  numpages  = {16},
  doi       = {10.1145/3586183.3606773},
  url       = {https://doi.org/10.1145/3586183.3606773}
}

@article{hindus1993capturing,
  author    = {Hindus, Debby and Schmandt, Chris and Horner, Chris},
  title     = {Capturing, structuring, and representing ubiquitous audio},
  journal   = {ACM Transactions on Information Systems},
  volume    = {11},
  number    = {4},
  year      = {1993},
  pages     = {376--400},
  numpages  = {25},
  publisher = {Association for Computing Machinery},
  address   = {New York, NY, USA},
  doi       = {10.1145/159764.159761},
  url       = {https://doi.org/10.1145/159764.159761}
}

@article{costley2021collaborative,
  author  = {Costley, Jamie and Fanguy, Mik},
  title   = {Collaborative note-taking affects cognitive load: The interplay of completeness and interaction},
  journal = {Educational Technology Research and Development},
  volume  = {69},
  number  = {2},
  year    = {2021},
  pages   = {655--671},
  doi     = {10.1007/s11423-021-09979-2},
  url     = {https://doi.org/10.1007/s11423-021-09979-2}
}

@misc{Zhang2024AiRNote,
  author = {{Deepgram Team}},
  title  = {Collaborative augmented reality note-taking with {AiRNote}},
  year   = {2024},
  url    = {https://deepgram.com/learn/ar-note-taking-airnote},
  note   = {Blog post. Accessed: 2026-01-08}
}

@article{kirsh2010thinking,
  author  = {Kirsh, David},
  title   = {Thinking with external representations},
  journal = {AI \& Society},
  volume  = {25},
  number  = {4},
  year    = {2010},
  pages   = {441--454},
  doi     = {10.1007/s00146-010-0272-8},
  url     = {https://doi.org/10.1007/s00146-010-0272-8}
}

@article{zhang1994representations,
  author  = {Zhang, Jiajie and Norman, Donald A.},
  title   = {Representations in distributed cognitive tasks},
  journal = {Cognitive Science},
  volume  = {18},
  number  = {1},
  year    = {1994},
  pages   = {87--122},
  doi     = {10.1207/s15516709cog1801_3},
  url     = {https://doi.org/10.1207/s15516709cog1801_3}
}

@misc{dynamicland,
  author = {Victor, Bret},
  title  = {Dynamicland},
  year   = {2018},
  url    = {https://dynamicland.org},
  note   = {Website. Accessed: 2026-01-08}
}

@inproceedings{billinghurst1999collaborative,
  author    = {Billinghurst, Mark and Kato, Hirokazu},
  title     = {Collaborative mixed reality},
  booktitle = {Proceedings of the First International Symposium on Mixed Reality},
  year      = {1999},
  pages     = {261--284}
}

@article{yang2024fostering,
  author  = {Yang, Jifan and Bednarski, Steven and Bullock, Alison and Harrap, Robin and MacDonald, Zack and Moore, Andrew and Graham, T. C. Nicholas},
  title   = {Fostering the {AR} illusion: A study of how people interact with a shared artifact in collocated augmented reality},
  journal = {Frontiers in Virtual Reality},
  volume  = {5},
  year    = {2024},
  pages   = {1428765},
  doi     = {10.3389/frvir.2024.1428765},
  url     = {https://doi.org/10.3389/frvir.2024.1428765}
}

@inproceedings{luo2025documents,
  author    = {Luo, Weizhou and Ellenberg, Mats Ole and Satkowski, Marc and Dachselt, Raimund},
  title     = {Documents in your hands: Exploring interaction techniques for spatial arrangement of augmented reality documents},
  booktitle = {Proceedings of the 2025 CHI Conference on Human Factors in Computing Systems},
  series    = {CHI '25},
  year      = {2025},
  publisher = {Association for Computing Machinery},
  address   = {New York, NY, USA},
  location  = {Yokohama, Japan},
  pages     = {1--22},
  numpages  = {22},
  doi       = {10.1145/3706598.3713518},
  url       = {https://doi.org/10.1145/3706598.3713518}
}

@article{ens2021,
  author  = {Ens, Barrett and Lanir, Joel and Tang, Anthony and Bateman, Scott and Lee, Gun and Piumsomboon, Thammathip and Billinghurst, Mark},
  title   = {Revisiting collaboration through mixed reality: The evolution of groupware},
  journal = {International Journal of Human-Computer Studies},
  volume  = {131},
  year    = {2019},
  pages   = {81--98},
  doi     = {10.1016/j.ijhcs.2019.05.011},
  url     = {https://www.sciencedirect.com/science/article/pii/S1071581919300606}
}

@inproceedings{rico2010usable,
  author    = {Rico, Julie and Brewster, Stephen},
  title     = {Usable gestures for mobile interfaces: Evaluating social acceptability},
  booktitle = {Proceedings of the SIGCHI Conference on Human Factors in Computing Systems},
  series    = {CHI '10},
  year      = {2010},
  publisher = {Association for Computing Machinery},
  address   = {New York, NY, USA},
  location  = {Atlanta, Georgia, USA},
  pages     = {887--896},
  doi       = {10.1145/1753326.1753458},
  url       = {https://doi.org/10.1145/1753326.1753458}
}

@inproceedings{fischelhalskovstickynotes,
  author    = {Fischel, Aron D. and Halskov, Kim},
  title     = {A survey of the usage of sticky notes},
  booktitle = {Extended Abstracts of the 2018 CHI Conference on Human Factors in Computing Systems},
  series    = {CHI EA '18},
  year      = {2018},
  publisher = {Association for Computing Machinery},
  address   = {New York, NY, USA},
  location  = {Montr{\'e}al, QC, Canada},
  pages     = {1--6},
  numpages  = {6},
  doi       = {10.1145/3170427.3188526},
  url       = {https://doi.org/10.1145/3170427.3188526}
}
\clearpage
\appendix
\section{Appendix}

\subsection{Introductory Participant Questions}
\label{appendixIntro}

\begin{enumerate}
  \item What is your name, age, and academic background?
  \item How often do you participate in group brainstorming or collaborative work?
  \item What tools do you typically use for collaborative tasks (e.g., sticky notes, whiteboards, digital tools)?
  \item What prior experience do you have with augmented or virtual reality, if any?
\end{enumerate}

\subsection{Post-Task Survey}
\label{appendixSurvey}

Participants completed a post-task survey after each condition. In Phase 2, returning participants only completed the AR portion of the survey. All questions used a 7-point Likert scale unless otherwise noted.

\subsection*{Participant Information}
\begin{itemize}
  \item Session ID
  \item Participant ID
\end{itemize}

\subsection*{Analog Condition: Paper Sticky Notes}

\begin{enumerate}
  \item How easy was it to keep track of ideas in this setup?
  \begin{itemize}
    \item Very difficult \dotfill 1 \dotfill 2 \dotfill 3 \dotfill 4 \dotfill 5 \dotfill 6 \dotfill 7 \dotfill Very easy
  \end{itemize}

  \item How smoothly did your conversation flow during this session?
  \begin{itemize}
    \item Not smooth at all \dotfill 1 \dotfill 2 \dotfill 3 \dotfill 4 \dotfill 5 \dotfill 6 \dotfill 7 \dotfill Very smooth
  \end{itemize}

  \item How often did the tools (pencil and sticky notes) interrupt or slow your discussion?
  \begin{itemize}
    \item Not at all \dotfill 1 \dotfill 2 \dotfill 3 \dotfill 4 \dotfill 5 \dotfill 6 \dotfill 7 \dotfill Very often
  \end{itemize}

  \item How connected did you feel to your partner (eye contact, shared focus)?
  \begin{itemize}
    \item Not at all \dotfill 1 \dotfill 2 \dotfill 3 \dotfill 4 \dotfill 5 \dotfill 6 \dotfill 7 \dotfill Very much
  \end{itemize}

  \item How mentally effortful did this setup feel?
  \begin{itemize}
    \item Not at all effortful \dotfill 1 \dotfill 2 \dotfill 3 \dotfill 4 \dotfill 5 \dotfill 6 \dotfill 7 \dotfill Very effortful
  \end{itemize}

  \item How easy was it to organize or group ideas?
  \begin{itemize}
    \item Very difficult \dotfill 1 \dotfill 2 \dotfill 3 \dotfill 4 \dotfill 5 \dotfill 6 \dotfill 7 \dotfill Very easy
  \end{itemize}

  \item How balanced did the participation feel between you and your partner?
  \begin{itemize}
    \item Not balanced at all \dotfill 1 \dotfill 2 \dotfill 3 \dotfill 4 \dotfill 5 \dotfill 6 \dotfill 7 \dotfill Very balanced
  \end{itemize}

  \item Did you coordinate actions with your partner (e.g., when to write, when to speak)?
  \begin{itemize}
    \item Never \dotfill 1 \dotfill 2 \dotfill 3 \dotfill 4 \dotfill 5 \dotfill 6 \dotfill 7 \dotfill Quite often
  \end{itemize}
\end{enumerate}

\subsection*{AR Condition: AnchorNote}

\begin{enumerate}
  \item How easy was it to capture ideas using the AR system?
  \begin{itemize}
    \item Very difficult \dotfill 1 \dotfill 2 \dotfill 3 \dotfill 4 \dotfill 5 \dotfill 6 \dotfill 7 \dotfill Very easy
  \end{itemize}

  \item How smoothly did the conversation flow?
  \begin{itemize}
    \item Not smooth at all \dotfill 1 \dotfill 2 \dotfill 3 \dotfill 4 \dotfill 5 \dotfill 6 \dotfill 7 \dotfill Very smooth
  \end{itemize}

  \item How often did AR-related issues (gesture problems, transcription errors, lag, alignment issues) interrupt or slow the discussion?
  \begin{itemize}
    \item Not at all \dotfill 1 \dotfill 2 \dotfill 3 \dotfill 4 \dotfill 5 \dotfill 6 \dotfill 7 \dotfill Very often
  \end{itemize}

  \item How connected did you feel to your partner while using AR?
  \begin{itemize}
    \item Not at all \dotfill 1 \dotfill 2 \dotfill 3 \dotfill 4 \dotfill 5 \dotfill 6 \dotfill 7 \dotfill Very connected
  \end{itemize}

  \item How mentally effortful did the AR interface feel?
  \begin{itemize}
    \item Not at all effortful \dotfill 1 \dotfill 2 \dotfill 3 \dotfill 4 \dotfill 5 \dotfill 6 \dotfill 7 \dotfill Very effortful
  \end{itemize}

  \item How intuitive was the method for creating sticky notes?
  \begin{itemize}
    \item Not at all intuitive \dotfill 1 \dotfill 2 \dotfill 3 \dotfill 4 \dotfill 5 \dotfill 6 \dotfill 7 \dotfill Very intuitive
  \end{itemize}

  \item How useful was the spatial placement of AR notes for understanding or remembering ideas?
  \begin{itemize}
    \item Not useful at all \dotfill 1 \dotfill 2 \dotfill 3 \dotfill 4 \dotfill 5 \dotfill 6 \dotfill 7 \dotfill Very useful
  \end{itemize}

  \item How balanced did the participation feel between you and your partner?
  \begin{itemize}
    \item Not balanced at all \dotfill 1 \dotfill 2 \dotfill 3 \dotfill 4 \dotfill 5 \dotfill 6 \dotfill 7 \dotfill Very balanced
  \end{itemize}

  \item Did you coordinate actions with your partner (e.g., when to write, when to speak)?
  \begin{itemize}
    \item Never \dotfill 1 \dotfill 2 \dotfill 3 \dotfill 4 \dotfill 5 \dotfill 6 \dotfill 7 \dotfill Very often
  \end{itemize}
\end{enumerate}

\subsection{Post-Task Semi-Structured Interview Guide}
\label{appendixInterview}

\textit{This interview lasted approximately 30 minutes. We followed a semi-structured format and asked follow-up questions or skipped questions as appropriate based on participant engagement and the quality of responses.}

\subsection*{Analog Session (Paper Sticky Notes and Whiteboard)}

\begin{enumerate}
  \item Walk me through your process during the analog session. What steps did you take to capture and organize ideas?
  \item Were there any moments when ideas were lost, forgotten, or repeated?
  \item Did the need to write or reference notes create pauses, waiting periods, or moments where you hesitated to speak?
  \item How did the tools make you feel during the discussion (e.g., comfortable, frustrated, distracted)?
  \item Where did you find yourself looking most often: at your partner, the notes, the whiteboard, or somewhere else?
  \item How did you and your partner divide roles during the session, if at all?
\end{enumerate}

\subsection*{AR Session (AnchorNote)}

\begin{enumerate}
  \item Walk me through your workflow during the AR session. How did you capture and manage ideas?
  \item Did AR prevent idea loss, or did it introduce new ways that ideas could be missed?
  \item Did AR-related breakdowns (e.g., gesture failures, transcription errors, lag) disrupt the flow of discussion?
  \item How did the AR system make you feel during the session (e.g., engaged, overwhelmed, distracted)?
  \item Did the AR system influence turn-taking or contribution patterns?
  \item Where did your attention go most often during the AR session?
  \item How easy or difficult was it to cluster, reorganize, or reference notes in AR?
  \item Did transcription quality affect your comfort or confidence when contributing ideas?
  \item Were there moments when AR enhanced or hindered shared understanding?
\end{enumerate}

\subsection*{Comparing Analog and AR Conditions}

\begin{enumerate}
  \item How did your process differ between the analog and AR conditions?
  \item Did either condition feel more awkward, stressful, smooth, or natural? Why?
  \item Which condition led to more pauses, hesitation, or waiting?
  \item Which condition led to more idea loss or forgetting, if any?
  \item Did either condition change when or how you contributed?
  \item Did either condition change the order in which ideas were externalized or developed?
\end{enumerate}

\subsection*{Advantages and Disadvantages}

\begin{enumerate}
  \item What were the biggest advantages of the analog tools?
  \item What were the biggest advantages of the AR tools?
  \item What were the main disadvantages or limitations of each setup?
  \item Which setup made collaboration feel more comfortable, confident, or connected?
  \item In what situations would you prefer analog tools versus AR tools?
\end{enumerate}

\subsection*{Viability of AR}

\begin{enumerate}
  \item Do you see AR as a replacement, complement, or specialized tool relative to analog sticky notes?
  \item What would AR need to do to match or exceed the analog brainstorming experience?
\end{enumerate}

\subsection*{Open Feedback}

\begin{enumerate}
  \item If you could redesign AnchorNote, what would you change?
  \item How might AR better support real-world group collaboration in the future?
  \item Is there anything else you would like to share about your experience?
\end{enumerate}

\subsection{Summarization Prompt}
\label{appendixPrompt}
The prompt sent to the LLM with the transcription to generate the concise summary that appears on the sticky notes is: \textit{"Create a 2-3 word title for this text. Only return the title, nothing else."}

\subsection{Implementation Details and Platform Components}
\label{appendixImplementation}

\Sys{} runs on Snap Spectacles and Lens Studio. We implemented explicit button-based interactions using SpectaclesUIKit components, while gesture detection and object manipulation relied on SpectaclesInteractionKit. Lens Studio's Automatic Speech Recognition (ASR) module handled real-time speech transcription.

Spectacles Sync Kit (SSK) supported multi-user synchronization by synchronizing instantiation of notes, real-time transform updates (position, rotation, and scale), and ownership-based deletion across participants.

To transform spoken ideas into concise, readable sticky notes, the system performed speech summarization using OpenAI’s ChatGPT API, and we accessed it via Lens Studio’s \texttt{RemoteServiceModule}. We selected a remote large language model rather than a fully local summarization approach because participants’ spoken contributions were often conversational, fragmented, and context-dependent, requiring abstraction and semantic compression beyond keyword extraction or rule-based methods. Preliminary tests with local, keyword-based summarization produced incomplete or overly literal notes that participants found less useful for later organization and reflection.

However, reliance on a remote service introduces latency and potential failures due to network connectivity or API availability. To preserve interaction continuity during collaborative sessions, we implemented a lightweight local keyword-based fallback that generates a provisional note when the remote service fails or times out. This fallback ensured that note creation remained responsive and predictable, while still enabling the system to generate high-quality summaries whenever the remote service was available.

We used custom scripting to synchronize text content during transcription and after summarization, and to enforce interaction constraints such as creator-only deletion and intentional long-pinch gestures.

\subsection{Quantitative Results}
\label{appendixQuantTable}

\begin{table}[H]
\centering
\caption{Mean (M) and standard deviation (SD in parentheses) for key measures across conditions (1--7 Likert scale). Higher scores indicate more of the measured attribute; for effort and interruptions, higher scores reflect greater burden. Phase 1 includes $N=12$ participants; Phase 2 includes $N=8$ returning participants. Each row corresponds to a single survey question, and we label constructs using the question’s conceptual focus (e.g., “mental effort,” “partner connection”).}
\begin{tabular}{cccc}
\toprule
Measure & Analog & Phase 1 & Phase 2 \\
\midrule
Mental effort & 2.92 (1.38) & 4.17 (1.64) & 3.63 (1.41) \\
Conversational smoothness & 6.00 (1.13) & 4.08 (1.16) & 4.88 (1.13) \\
Interruptions & 2.67 (1.61) & 5.42 (1.38) & 3.75 (0.89) \\
Partner connection & 5.50 (1.45) & 4.33 (1.15) & 4.38 (1.30) \\
Participation balance & 6.08 (1.08) & 4.58 (1.16) & 5.38 (1.30) \\
Spatial organization / usefulness & 5.92 (1.24) & 5.08 (1.56) & 5.75 (1.16) \\
\bottomrule
\end{tabular}
\end{table}

\subsection{Supplementary Qualitative Findings}
\label{appendixQual}

\subsubsection{Learning Curve and Familiarity Effects}
Returning participants in Phase~2 described a noticeably shorter learning curve compared to their initial exposure in Phase~1. Familiarity with the interaction flow reduced hesitation and monitoring behavior, allowing participants to engage more fluidly with the brainstorming task. P3 reflected that \textit{"getting used to it [would make] it faster to use,"} suggesting that novelty may explain some of the cognitive overhead observed in early sessions rather than inherent interaction costs. These observations support the interpretation that interaction predictability and learnability play a significant role in whether AR note-taking systems feel like a foreground interface or a background support during collaboration.

\subsubsection{Turn-Taking and Implicit Interaction Roles}
\label{appendixSpeakers}
Some participants observed that the act of creating notes implicitly structured conversational roles. During note creation, one participant often assumed a temporary \textit{"speaker"} role while the other became a \textit{"listener"} or observer. As P4 described, this dynamic occasionally emerged even when participants did not explicitly coordinate roles. While these interaction patterns did not prevent collaboration, they influenced conversational rhythm and turn-taking. Participants reported fewer such effects in Phase~2 because explicit controls made note creation feel more intentional and less disruptive.

\end{document}